\documentclass[runningheads]{svmult}

\usepackage{makeidx}   
\usepackage{graphicx}  
\usepackage{subeqnar}  
\usepackage{multicol}  
\usepackage{physprbb}  
\makeindex             

\begin{document}
\title*{The Classification of T Dwarfs}
\toctitle{The Classification of T Dwarfs}
%
%
\titlerunning{T Dwarf Classification}
%
\author{Adam J.\ Burgasser\inst{1}
\and J.\ Davy Kirkpatrick\inst{2}
\and Michael E.\ Brown\inst{1}}
\authorrunning{Adam J.\ Burgasser, et al.}
%
%
\institute{California Institute of Technology, MSC 103-33, 
Pasadena CA 91125, USA
\and Infrared Processing and Analysis Center, MSC 100-22,
Pasadena CA 91125, USA}

\maketitle              

{\noindent}{\em \lq\lq...one is forced to wonder where it will lead to, if
everyone who works on stellar spectra also introduces a new 
classification...''}

{\em -- Nils Duner (1899)}

\begin{abstract}
We discuss methods for classifying T dwarfs based on
spectral morphological features and indices.  T dwarfs are brown dwarfs 
which exhibit methane absorption bands at 1.6 and 2.2 ${\mu}m$.  
Spectra at red optical (6300--10100 {\AA})
and near-infrared (1--2.5 ${\mu}m$) wavelengths are presented,
and differences between objects are noted and discussed.  Spectral 
indices useful for classification schemes are presented.
We conclude that
near-infrared spectral classification is generally preferable 
for these cool objects,
with data sufficient to resolve the 1.17 and
1.25 ${\mu}m$ K I doublets
lines being most valuable.  
Spectral features sensitive to gravity are discussed, with the strength of the 
K-band peak used as an example.  Such features may be used to derive
a two-dimensional scheme based on temperature and mass, in analogy to the
MK temperature and luminosity classes.
\end{abstract}

\section{Introduction}
In the last decade, we have witnessed a flood of low-mass
star and brown dwarf detections, due to improvements in infrared array
technologies and the advent of large-scale, near-infrared sky surveys using
these detectors.  In 1995, the first brown dwarfs were discovered after
decades of failed searches -- the companion object Gl 229B \cite{nak95} 
and the
Pleiades brown dwarf Teide 1 \cite{reb95}. 
Since then, nearly one hundred brown dwarfs have been detected 
as companions to nearby stars and in the field, most 
made by new sky surveys such as the Two
Micron All Sky Survey (2MASS \cite{skr97}), the Sloan Digital Sky Survey
(SDSS \cite{yor00}), and the Deep Near-Infrared Survey (DENIS \cite{epc94}).
Others have been identified photometrically in various cluster surveys
(see contribution by R.\ Rebolo, these proceedings).
These surveys have finally brought brown dwarfs out of the realm of the 
theoretical and into that of the observational.

Two new spectral classes have also been defined as a result of 
these low-mass discoveries.  The {\em L spectral class}, 
comprised of objects
cooler than M9.5V, is
defined by \cite{kir99} and is 
discussed in the contribution by J.D.\ Kirkpatrick in these
proceedings.  The {\em T spectral class}, defined by the 
presence of CH$_4$ absorption at 2.2 ${\mu}m$ \cite{kir99}, 
is discussed here.
The prototype for this class, Gl 229B, shows CH$_4$ and H$_2$O
absorption bands throughout the near-infrared \cite{opp95,opp98},
exhibiting a spectral morphology similar to those of the
giant planets \cite{geb96}.
Gl 229B remained the solitary 
T dwarf until mid-1999, when 2MASS, SDSS,
and the New Technology Telescope Deep Field uncovered field analogs.
To date, at least 23 T dwarfs have been identified 
\cite{me99,me00a,me00c,cub99,tsv00,leg00}, a number sufficient to
begin consideration of spectral classification.

\subsection{Spectral Classification of Cool Dwarfs}

The classification of late-type dwarfs (which we define here as objects
later than spectral type K7V) has been done for over one hundred years
using a variety of features in photographic,
red optical, and near-infrared spectra.  Figure 1 diagrams
the wavelength and spectral type ranges of
a few major classification schemes related to the widely accepted
MK process.  
The classification of M dwarfs began with the conversion of
Secchi's type III stars \cite{sec1866} to class M 
by W.\ Fleming in the Henry Draper catalog
\cite{pic1890}.  This class was further divided by A.J. Cannon
into subtypes Ma, Mb, Mc, and 
Md, based on their spectral morphologies in the photographic regime
\cite{can01}.  Work by several authors using different instrumental
configurations extended M dwarf classification to decimal
systems up to type M6 
\cite{hof37,kui38,kui42,mor43,joh53}, based
primarily 
on the strength of TiO bands as temperature discriminants.    
Classification up to type M9V was done by 
P.C.\ Boeshaar in the photographic \cite{boe76,boe85} and 
J.D.\ Kirkpatrick in the red optical \cite{kir91,kir97}, based
on the appearance of TiO, VO, and CaOH bands.  The latter scheme
also makes use of spectral slope from 6300--9000 {\AA} as an additional
temperature discriminant.  Following the discovery of numerous objects
later than type M9.5V, J.D.\ Kirkpatrick defined and classified
the L spectral class in the red optical
based on the appearance of weakening oxides and strengthening 
metal hydride bands and
alkali lines \cite{kir99}.  
Classification
of the L class in the near-infrared is currently under investigation by
various authors.  
The MK(K) system \cite{mor43,joh53} makes use of 
additional spectral
features (e.g., CaOH, Na D lines) as luminosity discriminants 
in order to create a
two-dimensional scheme.
Luminosity class distinctions
have not been made for L dwarfs, as the coolest giant stars retain
strong TiO and VO bands \cite{kir99}. 

\begin{figure}[b]
\begin{center}
\includegraphics[width=0.6\textwidth,angle=90]{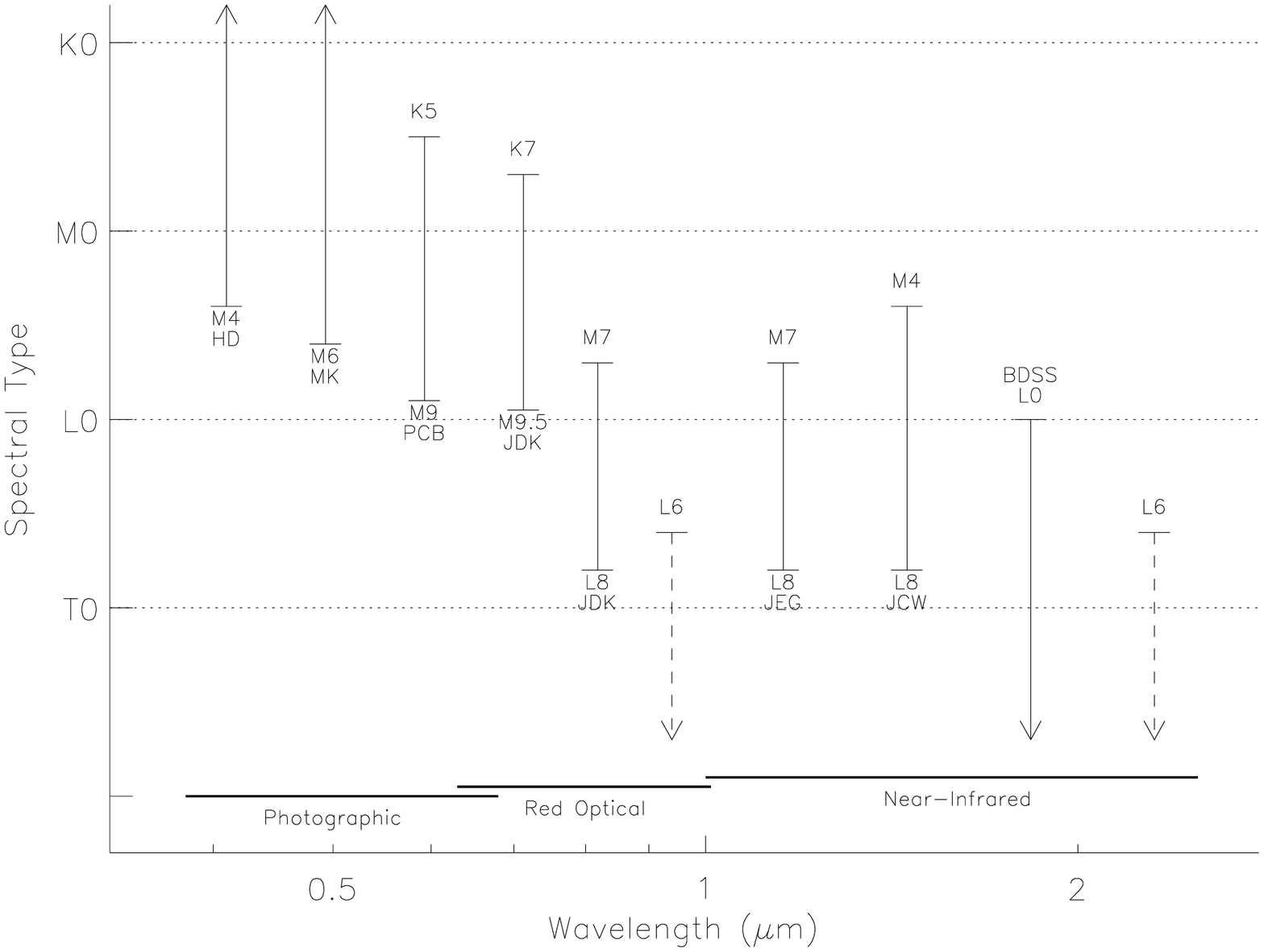}
\end{center}
\caption[]{Spectral typing schemes for M, L and T dwarfs over
the past century. 
The HD scheme is that of the Henry Draper catalog
\cite{pic1890,can01,hof37}; the MK (Yerkes) scheme is described in
\cite{kui38,kui42,mor43,joh53}; 
PCB are the P.C.\ Boeshaar schemes of 
\cite{boe76,boe85}; JDK are the M and L dwarf schemes of J.D.\
Kirkpatrick \cite{kir91,kir97,kir99}; near-infrared schemes for
L dwarfs are for J.E.\ Gizis (priv.\ comm.), J.C.\ Wilson (priv.\
comm.), and the NIRSPEC Brown Dwarf Spectroscopic Survey (I.\ McLean; priv.\
comm.); the dashed lines represent spectral schemes discussed
in the text, also described in \cite{me01}.}
\label{fig1}
\end{figure}

The schemes for T dwarf classification shown as dashed lines 
in Fig.\ 1 are the focus of this
article, and are discussed more fully in \cite{me01}.  
In {\S}2 we address red optical classification, based
on a spectral sample of bright T dwarfs identified by 2MASS and SDSS. 
In {\S}3 we discuss near-infrared
classification of a larger sample, and discuss spectral features 
important for classification.  
In {\S}4 we describe spectral
features in the near-infrared which may depend on
specific gravity, and could therefore append a second
dimension to a temperature scheme, in analogy to the MK
luminosity classes.  We
summarize our discussion in {\S}5.

\section{Red Optical Classification}

L dwarfs are classified by \cite{kir99} using
6300--10100\ {\AA} red optical spectra at 9\ {\AA} resolution, obtained with
the
Low Resolution Imaging Spectrograph (LRIS \cite{oke95}) mounted on the
Keck 10m Telescope.
As shown in Fig.\ 6 of \cite{kir99}, this spectral range samples
most of the prominent molecular and atomic features in late-M and L dwarf
spectra, and a \lq\lq recipe'' for spectral classification 
from M7V to L8V is presented based
on measurements of spectral indices on selected standards.  
In order to make a direct comparison with this L dwarf scheme,
we have obtained
optical spectra for nine T dwarfs utilizing the same instrumental setup 
as \cite{kir99}.  Data acquisition and reduction
are described in \cite{me00b,me01}.  Reduced 
spectra are shown in Fig.\ 2 for 2MASS 0559-14 \cite{me00a}
and Gl 570D \cite{me00a}, along with data for Gl 229B \cite{opp98}
and L dwarfs 2MASS 1507-16 (L5V), 2MASS 0920+35 (L6.5V), and
2MASS 1632+19 (L8V) \cite{kir99,kir00}.
Spectra are displayed on a log scale and offset in order to 
highlight features.

\begin{figure}[b]
\begin{center}
\includegraphics[width=\textwidth,angle=180]{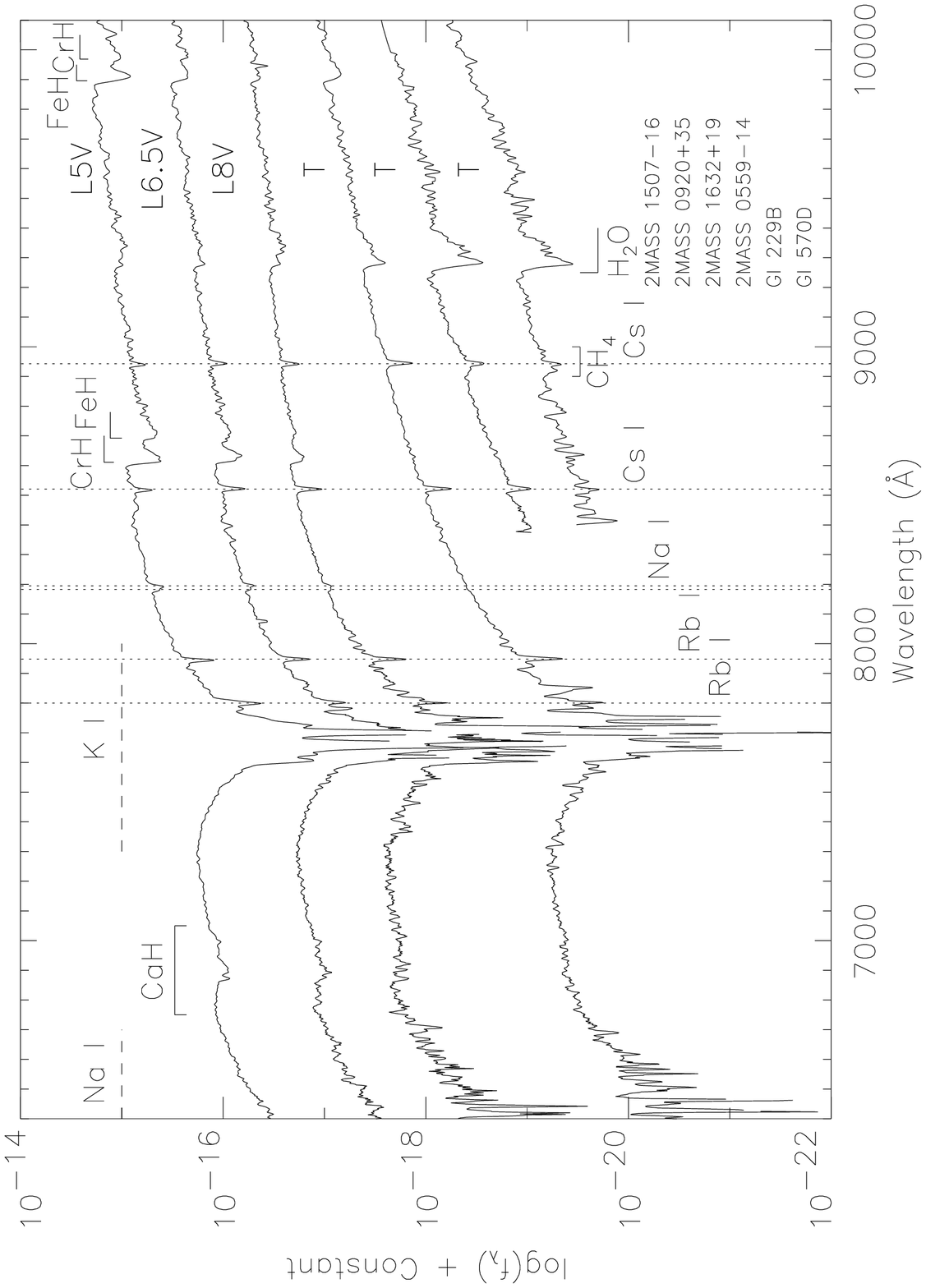}
\end{center}
\caption[]{Red optical spectra of L and T dwarfs.  Spectra are 
displayed on a log scale and offset by a constant factor.  
Major molecular and atomic features are noted.  L dwarf
spectra are from \cite{kir99,kir00}, Gl 229B data from 
\cite{opp98}.}
\label{fig2}
\end{figure}

A few general trends are readily apparent.  Most obvious is
the increased spectral slope from 8000--10100 {\AA} in the T dwarfs,
a feature which has been attributed to
pressure broadened K I at 7665 and 7699 {\AA} \cite{bur00,lie00}.  
A drop in flux shortward of 6600 {\AA} is due to the similarly broadened
Na I D lines at 5890 and 5895 {\AA} \cite{rei99}.
Increased H$_2$O
absorption at 9250 {\AA} is seen in the T dwarfs, strongest
in Gl 229B and Gl 570D.  
CrH (8611 {\AA}), FeH (8692 {\AA}),
and Na I 
(8183, 8195 {\AA}) features 
weaken in the late L dwarfs, and are completely absent
in the T dwarfs.  The behavior of the Cs I (8543, 8943 {\AA}) 
lines is less straightforward,
appearing to strengthen slightly from L5V to 2MASS 0559-14, but weakening
in Gl 229B and Gl 570D, with the 8943 {\AA} line becoming blended with
CH$_4$ at 8900 {\AA}.  
Rb I (7800, 7948 {\AA}) lines may be strengthening from mid-L to 
2MASS 0559-14, but lack of signal prevents their
detection in the other T dwarfs
(only five objects in our sample exhibit flux
shortward of 8000 {\AA}).
FeH (9896 {\AA}) is known to weaken from mid- to late-L \cite{kir99},
and is barely detected in 2MASS 1632+19; however, 2MASS 0559-14
shows a clear band, while Gl 229B and Gl 570D do not.
CaH (6750 {\AA}) may be present in all of the spectra shown here,
but is exceedingly weak in the latest L dwarfs and 2MASS 0559-14.  

It is clear, then, that there are qualitative differences in the optical 
spectra of L and T dwarfs.  The T dwarfs themselves appear to 
be distinguishable, based on their spectral slope, the presence of 
9896 {\AA} FeH,
and the 9250 {\AA}
H$_2$O band strength.  To explore these differences quantitatively,
we have examined a set of spectral indices, as described in Table 1.
Index values are plotted in Fig.\ 3
for the five brightest T dwarfs in our sample and Gl 229B
(open circles), as well as late-M and L dwarf standards from 
\cite{kir99} (solid
circles).  T dwarfs are ordered by their 
H$_2$O index.  The deepening of the 9250 {\AA} H$_2$O band
in the T dwarfs is readily
apparent, as is the increased spectral slope measured by the Color-d 
index.  Weakening of the 8611 {\AA} CrH and
9896 {\AA} FeH bands from types L to T is similarly seen in their
respective indices.
The KI-a index, which measures the relative depth of the broadened
K I doublet (or alternately its breadth), peaks at around L8V, then
decreases through the T dwarfs; this is 
due either to lower signal-to-noise in the
latter objects or the formation of KCl at T$_{eff}$ $\sim$ 950 K 
\cite{lod99}.
Cs I (8543 {\AA}) shows a general strengthening 
but with significant scatter, reflecting
its observed ambiguous behavior.

\begin{table}
\caption{T Dwarf Red Optical Spectral Indices}
\begin{center}
\renewcommand{\arraystretch}{1.4}
\setlength\tabcolsep{5pt}
\begin{tabular}{clll}
\hline\noalign{\smallskip}
Index & Numerator ({\AA}) & Denominator ({\AA}) & Feature Measured \\
\noalign{\smallskip}
\hline
\noalign{\smallskip}
KI-a & 7100.0--7300.0 & 7600.0--7800.0 & 7665, 7679 {\AA} K I doublet \\
Cs-a & Ave. 8496.1--8506.1 and 8536.1-8546.1 & 8516.1--8526.1 & 8543 {\AA} Cs I line$^a$ \\
CrH-a & 8580.0--8600.0 & 8621.0--8641.0 & 8611 {\AA} CrH band$^a$ \\
H$_2$O & 9225.0--9250.0 & 9275.0--9300.0 & 9250 {\AA} H$_2$O band \\
FeH-b & 9863.0--9883.0 & 9908.0--9928.0 & 9896 {\AA} FeH band$^a$ \\
Color-d & 9675.0--9875.0 & 7350.0--7550.0 & spectral slope$^a$ \\
\hline
\noalign{\smallskip}
\end{tabular}
\end{center}
$^a$ Also used as indices for L dwarfs in \cite{kir99}. \\
\label{tab2}
\end{table}

\begin{figure}[b]
\begin{center}
\includegraphics[width=\textwidth,angle=180]{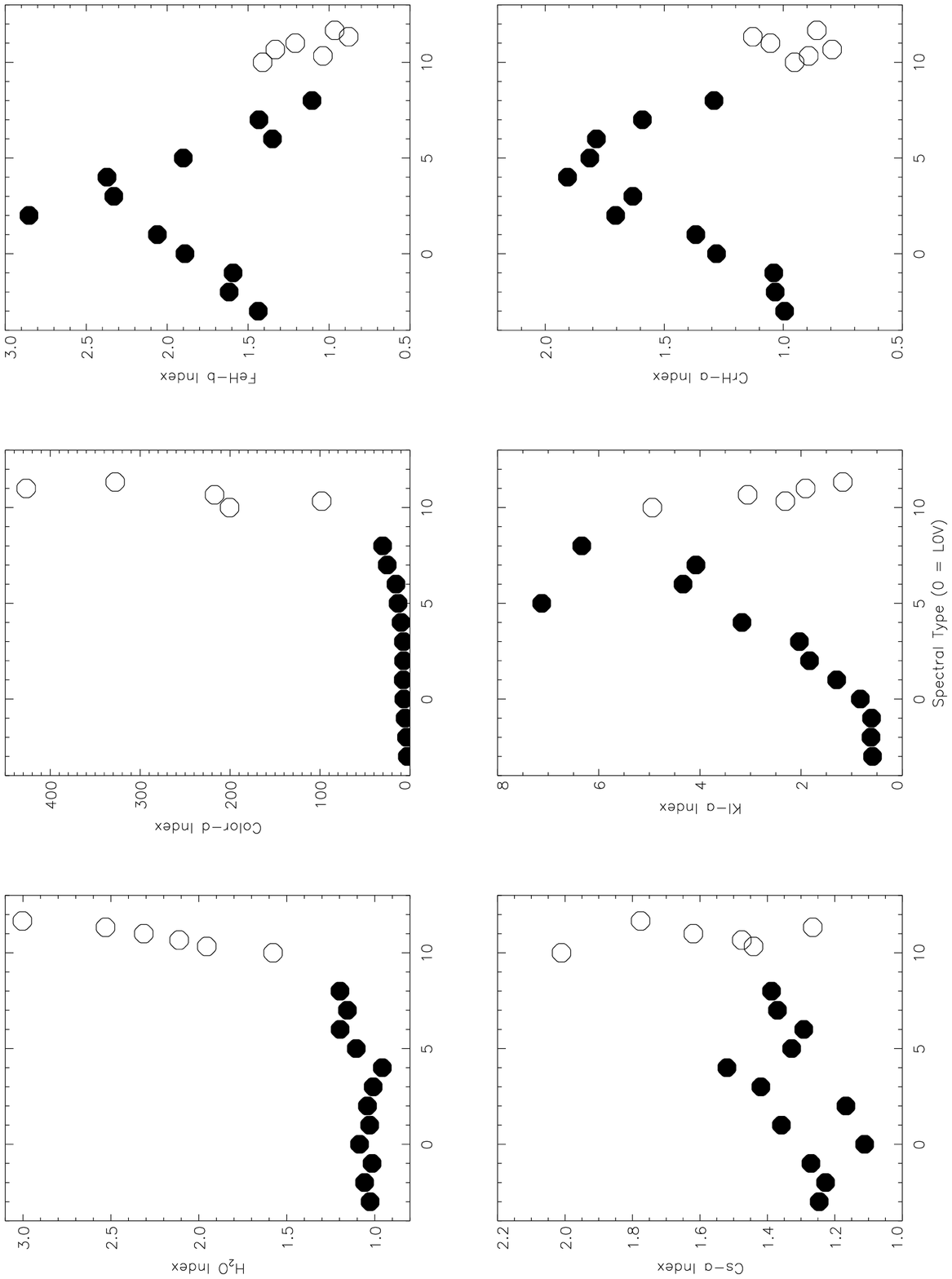}
\end{center}
\caption[]{Red optical spectral indices for late-M, L, and T dwarfs.
M and L dwarf values (filled circles)
are calculated using data from \cite{kir99,kir00}.
T dwarfs (open circles) are ordered by their H$_2$O index.
Indices are described in Table 1.}
\label{fig3}
\end{figure}

Based on these spectral indices and the quality of the observed spectra,
we make three conclusions regarding T dwarf classification in the optical:

\begin{itemize}
\item L and T dwarfs (later than 2MASS 0559-14) appear to be distinguishable
in the red optical, due to the absence of CrH (8611 {\AA}) and
FeH (8692 {\AA}) bands, increased 9250 {\AA} H$_2$O absorption, and
increased spectral slope in the T dwarfs.
\item T dwarfs can be separated in the red optical based on the 
presence or absence of the 9896 {\AA} FeH band, 
9250 {\AA} H$_2$O band strength, and 
spectral slope.  The observed scatter, however, restricts us
to only distinguishing ``early'' T dwarfs (e.g., 2MASS 0559-14) from ``late''
ones (e.g., Gl 570D).
\item Despite these successes, the faintness of these objects in the
optical (M$_R$ $\sim$ 24.6 for Gl 229B \cite{gol98}) makes it necessary
to consider classification in the near-infrared.
\end{itemize}

\section{Near-Infrared Classification}

The spectral energy distribution of T dwarfs peaks at around 1.27 ${\mu}m$,
the center of the J-band window, due to the combined effects of temperature
and H$_2$O, CH$_4$, and collision-induced (CIA) H$_2$ absorption features.
The 1--2.5 ${\mu}m$ region also encompasses the defining 2.2 ${\mu}m$
CH$_4$ band.  In our program to identify these cool brown dwarfs, we
have acquired 1--2.5 ${\mu}m$ spectra of 14 
T dwarfs at a resolution of R $\sim$ 100 (100{\AA} at 1 ${\mu}m$)
using the 
Near-Infrared Camera (NIRC \cite{mat94}) mounted on the Keck 10m Telescope.
Instrumental setup and data reduction are described in \cite{me01}.
Figure 4 displays these spectra (thick lines),
along with NIRC data for Gl 229B \cite{opp98},
and SDSS T dwarf
data from the literature \cite{str99,leg00} (higher resolution thin lines).  
A NIRC spectrum of 2MASS 0920+35 (L6.5V) is
shown for comparison.  
All spectra are normalized at their J-band peaks.

\begin{figure}[b]
\begin{center}
\includegraphics[width=\textwidth,angle=180]{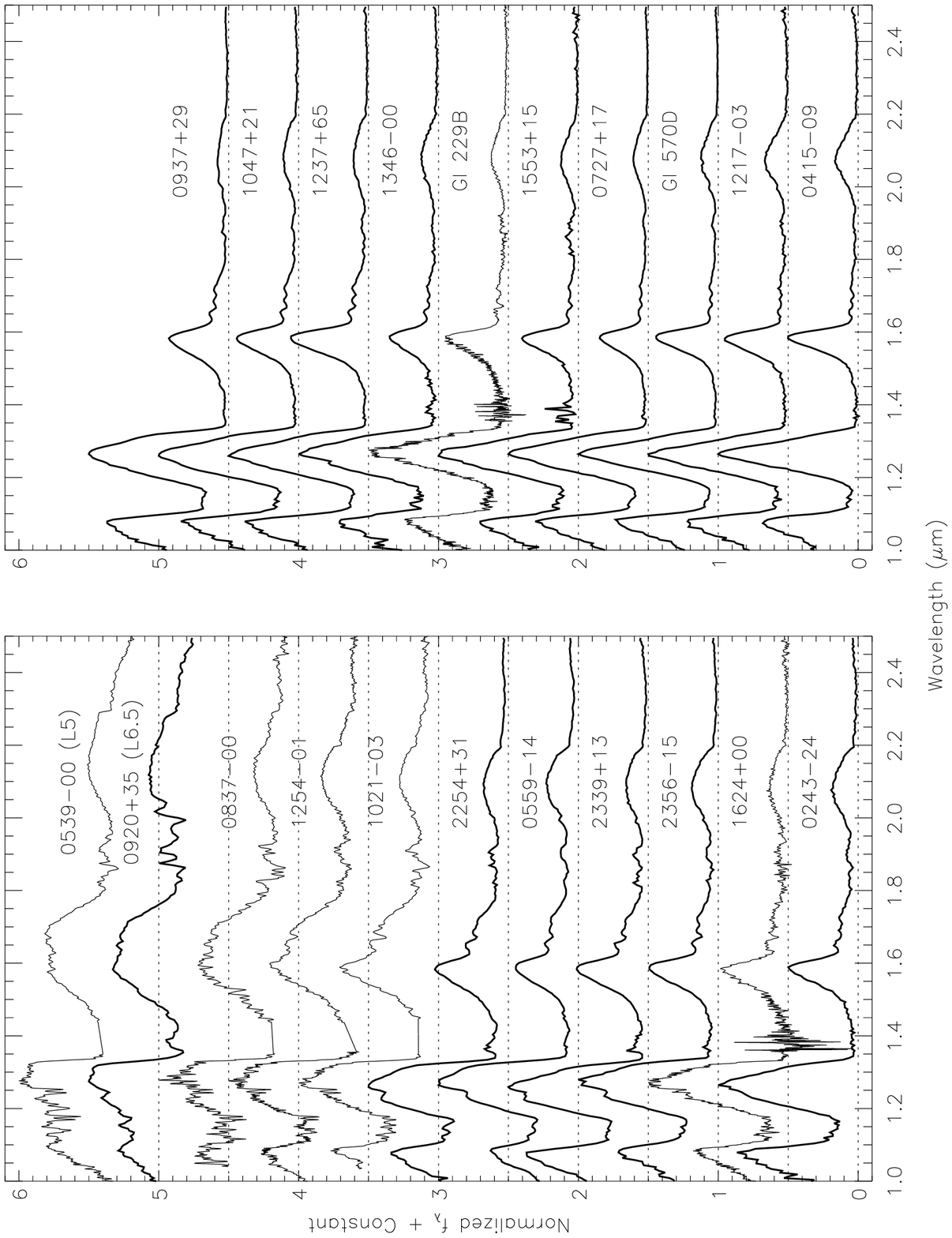}
\end{center}
\caption[]{Near-infrared T dwarf spectra.  Data obtained from NIRC
observations are shown in thick lines, while (higher resolution) 
data obtained from
the literature \cite{geb96,str99,leg00}
are shown as thinner lines.  Spectra are normalized at their J-band peaks
and offset by a constant.  Order is determined visually by comparison
of the depths of the 1.15 ${\mu}m$
H$_2$O/CH$_4$ and 1.6 ${\mu}m$ CH$_4$ features.}
\label{fig4}
\end{figure}

The spectra in Fig.\ 4 are ordered by visual inspection, based
on the increasing strengths of the 1.15 and 1.6 ${\mu}m$ absorption bands.
The 1.15 ${\mu}m$ feature begins in the late L dwarfs as a blend of
H$_2$O, two K I doublets (1.169 \& 1.177, 1.243 \& 1.252 ${\mu}m$), 
Na I (1.138, 1.141 ${\mu}m$), and
FeH (1.19, 1.21, 1.24 ${\mu}m$) \cite{mcl00}, but is later influenced
by CH$_4$ absorption at 1.1 ${\mu}m$
in the T dwarfs.  The deepening and broadening of this
feature and the adjacent H$_2$O/CH$_4$ absorption at 1.4 ${\mu}m$ causes
the J-band flux to become more peaked and narrow toward
the latest objects shown.  The 1.15 ${\mu}m$ feature also depresses the
1.243 \& 1.252 ${\mu}m$ K I doublet, unresolved in the NIRC data.  
There is significant suppression
of the H- and K-band peaks between SDSS 1021-03
and 2MASS 2254+31, coinciding with a large near-infrared color
difference between these objects (J-H $\sim$ 0.5 mag redder in SDSS 1021-03).  
This occurs
after CO at 2.3 ${\mu}m$, a prominent feature in late-M and L dwarfs,
disappears.  Toward later types, the K-band peak shape evolves from 
two bandheads at 2.2 (CH$_4$) and 2.3 ${\mu}m$ (CO), to a rounded
peak with a kink at 2.17 ${\mu}m$, to a symmetric peak at 2.07 ${\mu}m$ in
2MASS 0415-09.

\begin{figure}[b]
\begin{center}
\includegraphics[width=\textwidth,angle=180]{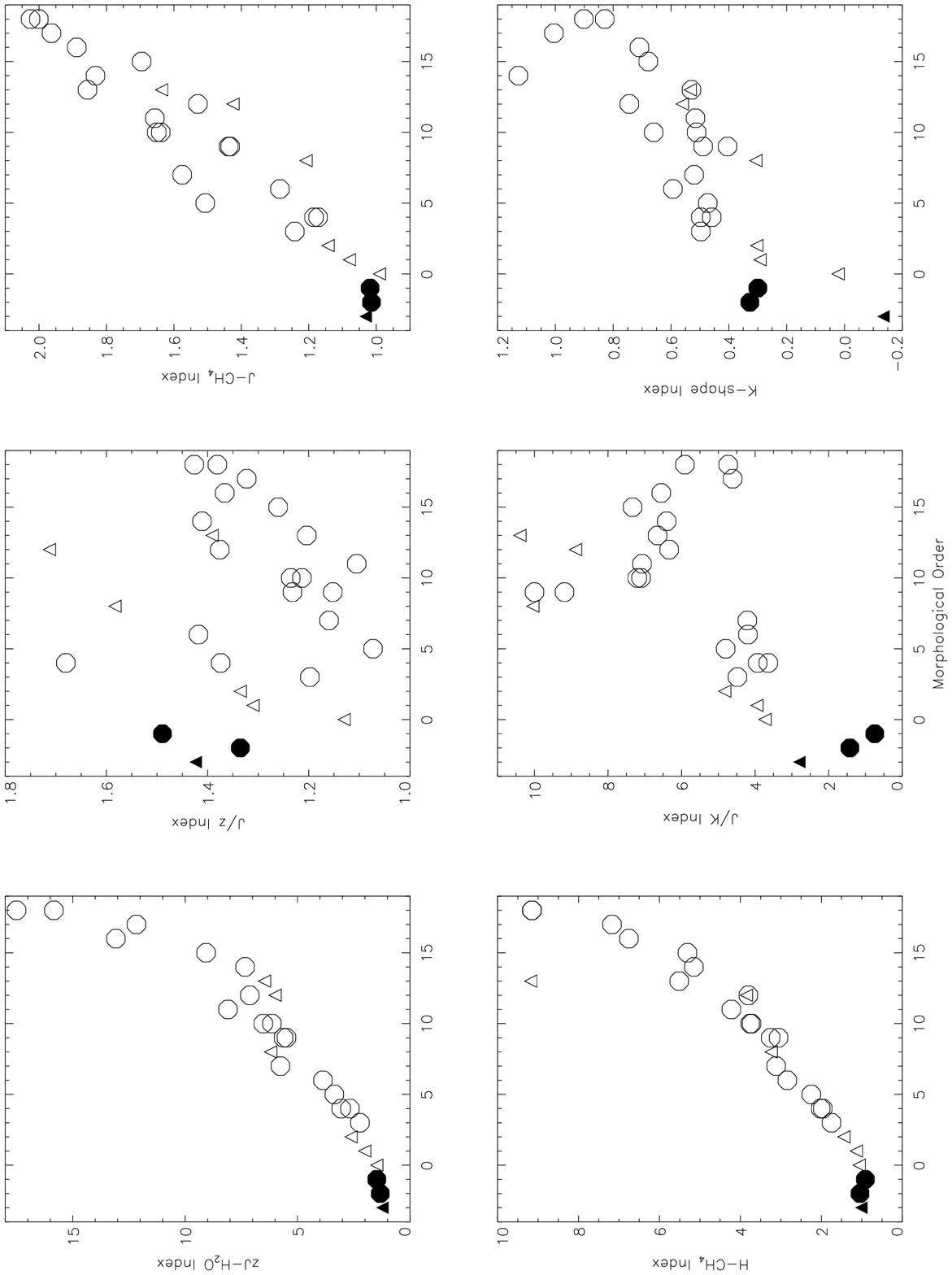}
\end{center}
\caption[]{Near-infrared spectral indices for L (filled symbols)
and T dwarfs (open symbols).  Measurements of NIRC data are shown
as circles, while data obtained from the literature 
\cite{geb96,str99,leg00} are plotted as triangles.  T dwarfs are
ordered as shown in Fig.\ 4.  Indices are described in Table 2.}
\label{fig5}
\end{figure}

In order to quantitatively explore these 
observed spectral variations, we have defined near-infrared spectral
indices in analogy to those derived for the 
red optical spectra.  These indices are
described in Table 3 and plotted in Fig.\ 5 for the NIRC data
(circles), along with values computed from
L and T dwarf spectral data from the literature \cite{geb96,str99,fan00,leg00} 
(triangles).
Filled and open symbols
denote L and T dwarfs, respectively.  T dwarfs are ordered as in
Fig.\ 4, while L dwarfs are ordered by their spectral type from \cite{kir99}.
Aligned symbols represent multiple spectra for the same object, 
allowing us 
to estimate the approximate noise in these values, as well as compare
spectra of different resolutions.
There is an obvious correlation between the depth of the 1.15 ${\mu}m$ 
H$_2$O/CH$_4$ band, the 1.3 ${\mu}m$ CH$_4$ band wing, and the 1.6 ${\mu}m$
CH$_4$ band, all of which steadily increase from the late-L dwarfs through
2MASS 0415-09.  The K-shape index (measuring the change of the 2.17 ${\mu}m$
kink) also increases toward unity (no kink) over this range, although with
more scatter.  Similarly, the J/K peak ratio appears to generally increase,
although there is less agreement between the NIRC data and data from the 
literature.  One object, 2MASS 0937+29 (see below) is particularly
discrepant.  Finally, the J/z peak ratio shows the least correlation with
morphological order, with variations no more than 40\% of the mean,
suggesting noise as a possible contributor.

Overall, their appears to be a good correlation in the strengths of the
CH$_4$ and H$_2$O absorption band strengths, and these are likely to be
crucial features for spectral typing in the near-infrared.  K I lines,
unresolved in this data, are also important indicators of temperature
(being higher energy lines than the broadened 7665 \& 7699 {\AA} K I doublet
in the red optical), so that classification spectra in this wavelength
regime
should be of sufficient resolution to permit measurement of these lines.
These data can be easily obtained using existing instrumentation.

\begin{table}
\caption{T Dwarf Near-Infrared Spectral Indices}
\begin{center}
\renewcommand{\arraystretch}{1.4}
\setlength\tabcolsep{5pt}
\begin{tabular}{llll}
\hline\noalign{\smallskip}
Index & Numerator (${\mu}m$) & Denominator (${\mu}m$) & Feature Measured \\
\noalign{\smallskip}
\hline
\noalign{\smallskip}
zJ-H$_2$O &  1.25--1.28 & 1.13--1.16 & 1.15 ${\mu}m$ H$_2$O/CH$_4$ band \\
J/z &  1.26--1.28 & 1.06--1.08 & J-band/z-band peak ratio \\
J-CH$_4$ & 1.26--1.275 & 1.295--1.31  & 1.3 ${\mu}m$ CH$_4$ absorption wing \\
H-CH$_4$ &  1.57--1.59 & 1.63--1.65 & 1.6 ${\mu}m$ CH$_4$ band \\
J/K &  1.26--1.28 & 2.06--2.09 & J-band/K-band peak ratio \\
K-shape &  2.10--2.11 $minus$ 2.16--2.17 & 2.17--2.18 $minus$ 2.19--2.20 & K-band shape (2.17 ${\mu}m$ kink) \\
\hline
\end{tabular}
\end{center}
\label{tab1}
\end{table}

\section{Additional Classification Parameters in the Near-Infrared}

The discrepancy seen in the J/K peak ratio for 2MASS 0937+29 is fairly
obvious in Fig.\ 5, and can be seen in 
Fig.\ 6a, which shows the 1.5--2.5 ${\mu}m$ spectra of 2MASS 0937+29 and 
2MASS 1047+21, both normalized at their J-band peaks. The similarity
of these objects
at H-band is in striking contrast to the notable suppression of
the K-band peak of
2MASS 0937+29.  This object is the bluest T dwarf so far identified, 
with J-K$_s$ = --0.89$\pm$0.24, as compared to the mean J-K$_s$ $\sim$ 0 for
all other 2MASS T dwarfs. 
The K-band peak is shaped by the combined 
effects of H$_2$O (1.8--2.1 ${\mu}m$), CH$_4$ (2.2-2.6 ${\mu}m$), and
CIA H$_2$ (centered around 2.5 ${\mu}m$).  The latter feature is 
strongest at K-band, and, being a collisional process, is highly
sensitive to the total photospheric pressure.  Zero-metallicity models
show that increased gravity (and therefore photospheric temperature)
results in bluer H--K colors due to increased H$_2$ absorption, 
an effect which becomes more
pronounced toward cooler T$_{eff}$ \cite{sau94}.  

To examine this effect,
we have computed spectral indices for models provided by D.\ Saumon
(priv.\ comm.).  Figure 6b shows the behavior of the J/K peak ratio as
compared to
1.6 ${\mu}m$ CH$_4$ band strength for gravities of 300, 1000,
and 3000 m s$^{-2}$.  Overplotted are index values measured for the
2MASS T dwarfs (open circles).  Gl 570D is separately indicated (solid
triangle); this companion 
object has an estimated T$_{eff}$ = 750$\pm$50 K and g = 800--2000 m s$^{-2}$
\cite{me00a}, in consensus with its location in Fig.\ 6b relative 
to the models.  
The large spread in J/K index values versus gravity for the
models is obvious, and is greater at lower
temperatures.
We see that 2MASS 0937+29 and 2MASS 1047+21 (solid circles) lie on opposite
sides of the model g = 3000 m s$^{-2}$ line, consistent with 
the former having a higher surface gravity.  
This example suggests that 
comparison of the K-band peaks for a sample of classified
T dwarfs may allow
segregation by gravity (and therefore mass), in analogy
to the dwarf/giant segregation made in the MK luminosity classes.
Variations in metallicity may also play an important role in H$_2$
strength, and the effects of CIA H$_2$ absorption at other wavelengths
(i.e., near peaks at 0.83 and 1.25 ${\mu}m$) needs to be similarly investigated.
Alkali lines, particularly the 1.17 and 1.25 ${\mu}m$ K I doublets, may also
elucidate gravity and/or metallicity effects, and provide a further
discriminant for a two-dimensional classification scheme.

\begin{figure}[b]
\begin{center}
\includegraphics[width=\textwidth,angle=180]{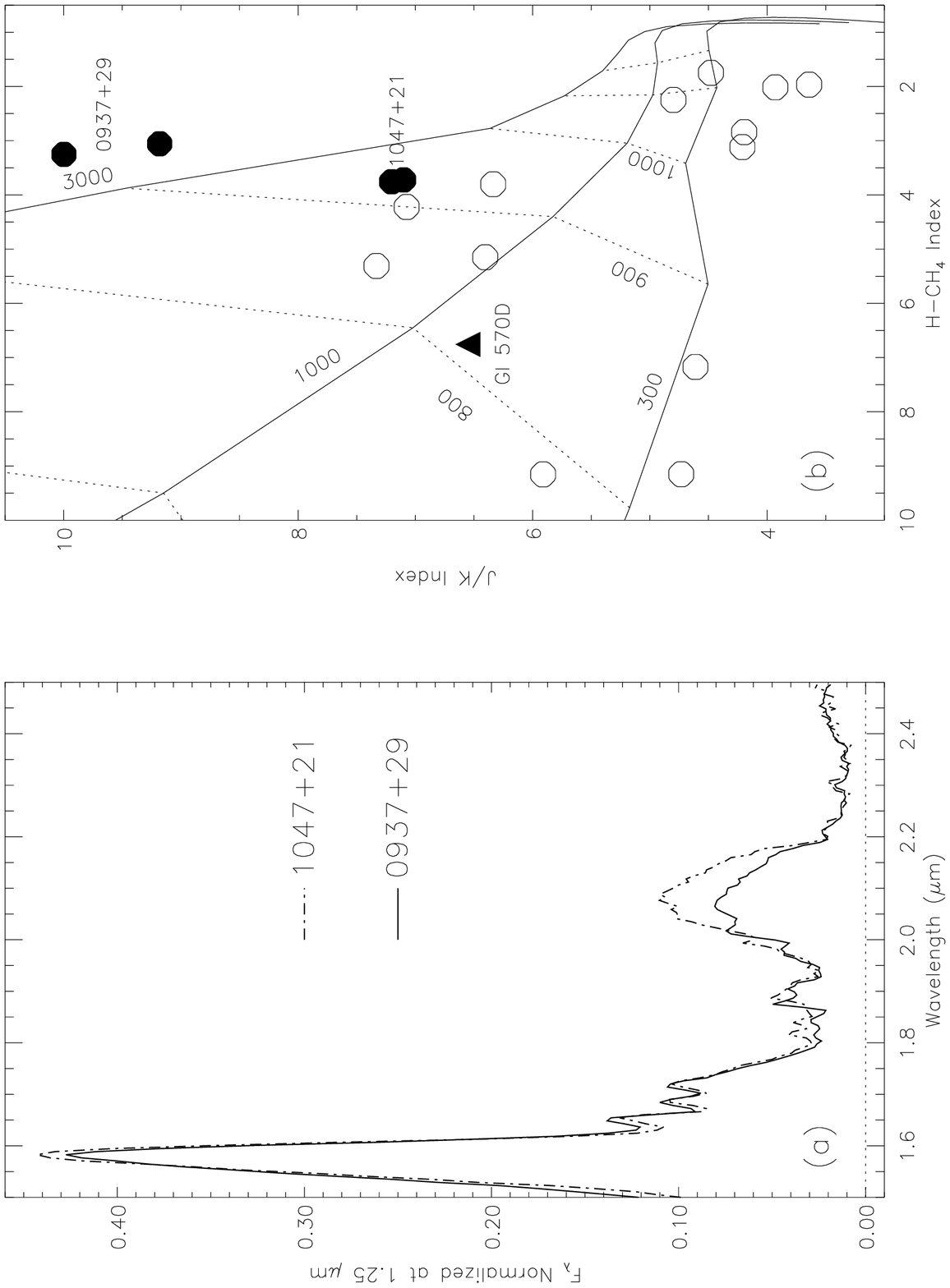}
\end{center}
\caption[]{Spectral estimation of T dwarf gravities. (a) Spectral
comparison between
2MASS 0937+29 (solid line) and 2MASS 1047+21 (dashed line); both
spectra are normalized at their J-band peaks. (b) H-CH$_4$ index
versus J/K peak ratio for 2MASS T dwarfs and T dwarf models provided
by D.\ Saumon (priv.\ comm.)  Lines trace out model values for 
gravities of 300, 1000, and 3000 m s$^{-2}$.}
\label{fig6}
\end{figure}

\section{Summary}

We have discussed the prospects for spectral typing T dwarfs in the 
red optical and near-infrared, based on analysis of spectral data obtained
by the authors and from the literature.  Basic morphological differences
are seen in both regimes, supporting the viability of considering
classification schemes for these cool brown dwarfs.  Several conclusions
may be drawn from the preceding discussion:

\begin{itemize}
\item The red 
optical spectra of T dwarfs is different than that of the L dwarfs,
due to differences in spectral slope, 9250 {\AA} H$_2$O band strength,
and the presence or absence of various hydride bands and alkali lines.
The faintness of the T dwarfs at these wavelengths,
however, favors classification schemes at longer
wavelengths.
\item A continuous sequence of T dwarfs can be seen in near-infrared
spectral data, both at low (R $\sim$ 100) and moderate (R $\sim$ 600)
resolutions.  This sequence is dominated by the bands of CO 
(in the earliest T dwarfs),
H$_2$O, CH$_4$, and CIA H$_2$, 
while the 1.25 ${\mu}m$ K I doublet, unresolved in the NIRC data, may
show a corresponding progression.  The clear variation in 
features and greater facility of obtaining reasonable signal-to-noise
spectra favors this wavelength regime for classification.
\item Spectral features that are sensitive to specific gravity 
may allow development of a two-dimensional scheme analogous
to the MK luminosity class, but instead based on object mass.  
The strength of the K-band peak appears
to be such a feature, due to pressure-sensitive CIA H$_2$ 
absorption.  Calibration of this effect and the influence of 
metallicity remains to be explored. 
\end{itemize}

A.J. Burgasser acknowledges S.K.\ Leggett, B.R.\ Oppenheimer, and M.A.\ 
Strauss for providing useful spectral data for comparison, and D.\ Saumon for
use of unpublished 
model spectra.  Enlightening discussions over the role of H$_2$ were made
with A.\ Burrows, M.\ Marley, and D.\ Saumon.  AJB also acknowledges the
contributions of the other members of the 2MASS Rare Objects Team: R. Cutri,
C. Dahn, J. Gizis, J. Liebert, B. Nelson, and I.N. Reid.
Data presented herein were obtained at the W.\ M.\ Keck Observatory
which is operated as a scientific partnership among the California Institute of
Technology, the University of California, and the National Aeronautics and Space
Administration.  The Observatory was made possible by generous financial 
support of the W.\ M.\ Keck Foundation.
This contribution makes use of data from the Two Micron
All Sky Survey, which is a joint project of the University of
Massachusetts and the Infrared Processing and Analysis Center,
funded by the National Aeronautics and Space Administration and
the National Science Foundation.

%

\end{document}